\documentclass[conference]{IEEEtran}
\usepackage{amsmath,amsfonts}
\usepackage{algorithm}

\usepackage{algpseudocode}

\usepackage{array}
\usepackage[caption=false,font=normalsize,labelfont=sf,textfont=sf]{subfig}
\usepackage{textcomp}
\usepackage{stfloats}
\usepackage{url}
\usepackage{tabularx}
\usepackage{verbatim}
\usepackage{graphicx}
\usepackage{cite}
\begin{document}
\title{DDRM: Distributed Drone Reputation Management for Trust and Reliability in Crowdsourced Drone Services}

\author{
    \IEEEauthorblockN{Junaid Akram\IEEEauthorrefmark{1}, Ali Anaissi\IEEEauthorrefmark{1}\IEEEauthorrefmark{2}}
    \IEEEauthorblockA{\IEEEauthorrefmark{1}School of Computer Science, The University of Sydney, Camperdown NSW 2008, Australia\\
    \IEEEauthorrefmark{2}TD School, University of Technology Sydney, Ultimo NSW 2007, Australia\\
    Email: jakr7229@uni.sydney.edu.au, ali.anaissi@sydney.edu.au}
}

\maketitle

\begin{abstract}
This study introduces the Distributed Drone Reputation Management (DDRM) framework, designed to fortify trust and authenticity within the Internet of Drone Things (IoDT) ecosystem. As drones increasingly play a pivotal role across diverse sectors, integrating crowdsourced drone services within the IoDT has emerged as a vital avenue for democratizing access to these services. A critical challenge, however, lies in ensuring the authenticity and reliability of drone service reviews. Leveraging the Ethereum blockchain, DDRM addresses this challenge by instituting a verifiable and transparent review mechanism. The framework innovates with a dual-token system, comprising the Service Review Authorization Token (SRAT) for facilitating review authorization and the Drone Reputation Enhancement Token (DRET) for rewarding and recognizing drones demonstrating consistent reliability. Comprehensive analysis within this paper showcases DDRM's resilience against various reputation frauds and underscores its operational effectiveness, particularly in enhancing the efficiency and reliability of drone services.
\end{abstract}


%
\IEEEpeerreviewmaketitle

\section{Introduction}

In the wake of the 21st century, technological advancements have reshaped the landscape of various sectors, with Unmanned Aerial Vehicles (UAVs), commonly known as drones, leading the charge\cite{akram2023chained, akram2022bc}. These innovations have expanded their reach from conventional applications such as surveillance\cite{munawar2022framework, munawar2022civil} and logistics\cite{munawar2023drone} to critical environmental monitoring efforts\cite{tahir2022automatic}, and now, into the domain of emergency management, specifically bushfire detection and management. The evolution of drone technology has been symbiotically linked with the concept of crowdsourced drone services within the burgeoning Internet of Drone Things (IoDT) ecosystem, heralding a new era of collaborative and decentralized environmental stewardship\cite{akram2024DroneSSL}.

Central to our discourse is the innovative utilization of crowdsourced drone services for bushfire management, a critical concern in many parts of the world. Our system introduces a pioneering marketplace that bridges the gap between bushfire management authorities and drone operators. In this marketplace, the authorities, acting as data consumers, can access indispensable information for bushfire detection and management, collected by drone operators who serve as data providers\cite{lakhdari2020composing}. This collaborative approach leverages the collective capability of drone technology enthusiasts, enabling the collection of vital data that significantly enhances the efficiency of bushfire detection and management strategies\cite{mukherjee2023isocialdrone, bahutair2021multi}.

This crowdsourced framework not only democratizes access to drone technology, making it feasible for entities of varying scales to contribute towards a common goal but also fosters an environment of innovation and economic growth within the realm of environmental management and disaster response\cite{akram2024DroneSSL, munawar2023drone}. By empowering drone operators to participate in critical data collection activities, our system facilitates the generation of essential statistics and insights necessary for timely and effective bushfire management\cite{munawar2022framework, 10535995}.

Moreover, the integrity and reliability of the data collected through this crowdsourced ecosystem are of paramount importance. Prior studies have highlighted the influence of reviews and feedback in shaping user engagement within technology ecosystems\cite{mukherjee2022disastdrone, chatterjee2001online}. In the context of our system, ensuring the authenticity of data and the credibility of drone operators becomes crucial, as the effectiveness of bushfire management efforts heavily relies on the accuracy and timeliness of the information provided\cite{islam2021blockchain, 10492460}. Addressing potential challenges related to data integrity and feedback manipulation, our framework proposes robust mechanisms to safeguard against fraudulent activities and ensure the genuineness of contributions from all participants\cite{mukherjee2020edgedrone,9422738, 10130620}.

In response to these considerations, the Distributed Drone Reputation Management (DDRM) framework, anchored in the Ethereum blockchain, plays a crucial role. DDRM enables a decentralized, peer-to-peer (P2P) mechanism for verifying the trustworthiness of drone service providers within the ecosystem. By eliminating central points of vulnerability and enhancing transparency, DDRM ensures that bushfire management authorities can rely on the data obtained through our marketplace for making informed decisions\cite{munawar2022civil, tahir2022automatic,10547221}.

As drone technology continues to advance, the DDRM framework is designed to be flexible and adaptable to these changes. The use of smart contracts allows for the seamless integration of new functionalities and protocols as they are developed. This adaptability ensures that DDRM remains relevant and effective in dynamic real-world scenarios. Additionally, the unique characteristics of crowdsourced drone services, such as their ability to access remote and hazardous areas, differentiate them from other crowdsourcing services and underscore the necessity of a specialized reputation management system like DDRM.
Furthermore, while DDRM is tailored for the IoDT ecosystem, its underlying principles can be applied to other scenarios requiring robust reputation management. The flexibility and scalability of DDRM make it a versatile solution for various applications beyond drone services. This comparative analysis with non-blockchain systems highlights DDRM's superior ability to provide a trustworthy and scalable reputation management solution.

This paper delves into the intricacies of leveraging crowdsourced drone services for bushfire management, elucidating the transformative potential of our system to not only enhance the efficacy of environmental monitoring and disaster response strategies but also to pioneer a new paradigm in the collaborative management of natural calamities.  The major contributions are outlined as follows:

\begin{itemize}
    \item We present a novel reputation management framework specifically designed for the IoDT ecosystem. DDRM leverages a Peer-to-Peer (P2P) consortium network of randomly selected endorser nodes, which is built upon the Ethereum blockchain. This innovative approach ensures the verifiability and authenticity of drone service reviews, addressing the critical need for trust and integrity in the feedback mechanism within the IoDT.
    
    \item A distinctive aspect of our contribution is the introduction of a two-token system, comprising the Service Review Authorization Token (SRAT) and the Drone Reputation Enhancement Token (DRET). The SRAT is instrumental in facilitating the authorization of reviews, ensuring that only legitimate and verified interactions contribute to the feedback system. Concurrently, the DRET serves to enhance the reputation of drone service providers who consistently demonstrate reliability and quality in their offerings. This dual-token approach not only incentivizes honest behavior among drone operators but also significantly elevates service credibility within the IoDT ecosystem.
    
    \item An extensive security and performance analysis of DDRM has been conducted. This study critically evaluates DDRM's resilience against potential reputation frauds, a prevalent concern in decentralized ecosystems. Additionally, we assess the operational effectiveness of DDRM within the IoDT environment, ensuring that our proposed framework meets the high standards required for real-world application. This evaluation provides valuable insights into the robustness of DDRM and its capacity to safeguard the integrity of drone service reviews.
    
\end{itemize}

The structure of this paper unfolds as follows. Section 2 sheds light on the foundational knowledge of the Ethereum blockchain and the ERC20 token standard and navigates through current literature of blockchain-based reputation systems, particularly focusing on drone networks. The intricacies of the DDRM framework, its components, and underlying algorithms are elucidated in Section 3. In Section 4, we critically analyze and evaluate the DDRM's robustness and security measures, and discuss potential limitations and challenges faced by our proposed framework. We encapsulate our findings and chart future research directions in Section 5.

\begin{table*}[h]
\centering
\caption{DDRM and Related Work Comparison Table.}
\label{table:comparison}
\begin{tabular}{|p{0.8cm}|l|p{2cm}|p{2cm}|p{1.8cm}|p{1cm}|p{1cm}|p{1cm}|p{1cm}|p{1cm}|}
\hline
\textbf{Paper} & \textbf{Model Type} & \textbf{Underlying Architecture} & \textbf{Primary Focus} & \textbf{Implementation} & \multicolumn{4}{c|}{\textbf{Attacks Prevented}} & \textbf{Gas Cost Analysis}\\
\cline{6-9}
 & & & & & \textbf{Sybil} & \textbf{Ballot Stuffing} & \textbf{Bad Mouthing} & \textbf{Collusion} & \\
\hline
\cite{ramachandiran1using} & Non-incentive & Bitcoin & Review Integrity & Not Done & Not Mitigated & Not Mitigated & Not Mitigated & Not Mitigated & N/A \\
\hline
\cite{dennis2016rep} & Non-incentive & Variation of Bitcoin Protocol & P2P reputation and scalability & Not Done & Mitigated & Not Mitigated & Not Mitigated & Mitigated & N/A \\
\hline
\cite{azad2018privbox} & Non-incentive & Homomorphic Cryptographic Non-Interactive Zero-Knowledge Proofs & Reviewer privacy in decentralized environment & Done & Not Mitigated & Not Mitigated & Mitigated & Not Mitigated & N/A \\
\hline
\cite{park2018smart} & Non-incentive & Ethereum Smart Contracts & P2P marketplace reputation & Not Done & Not Mitigated & Not Mitigated & Not Mitigated & Mitigated & Not Done \\
\hline
\cite{carboni2015feedback} & Incentive-based & Bitcoin with Vouchers & Feedback abuse control & Not Done & Partially Mitigated & Not Mitigated & Mitigated & Not Mitigated & N/A \\
\hline
\cite{schaub2016trustless} & Incentive-based & Blind Signatures & Token usage for reviews & Not Done & Mitigated & Not Mitigated & Not Mitigated & Not Mitigated & N/A \\
\hline
\cite{salah2019blockchain} & Incentive-based & Ethereum with IPFS & Automation of incentives & Done & Not Mitigated & Partially Mitigated & Not Mitigated & Not Mitigated & N/A \\
\hline
DDRM & Incentive-based & Ethereum, ERC20 Tokens, P2P Network & Tamper-proof, fraud resilience, review validation & Done & Mitigated & Mitigated & Mitigated & Mitigated & Done \\
\hline
\end{tabular}
\end{table*}

\section{Related Work}

The burgeoning domain of the Internet of Drone Things (IoDT) has underscored the imperative need for scalable and authentic review mechanisms, propelling the evolution of underlying technologies such as Ethereum's blockchain. Our DDRM framework leverages Ethereum and its ERC20 token standard to ensure the integrity and verifiability of drone service reviews, pivotal for fostering trust within the IoDT ecosystem.

\subsection{Ethereum and ERC20 Tokens}
Ethereum, a public blockchain renowned for its programmable feature via smart contracts, facilitates the development of decentralized applications (dApps) \cite{wood2014ethereum}. Unlike its predecessors, Ethereum allows for a high degree of customization through smart contracts, primarily coded in Solidity \cite{dannen2017introducing}. The essence of Ethereum lies in its ability to transcend traditional transaction protocols, enabling users to create bespoke protocols. This flexibility forms the backbone of our DDRM framework, utilizing ERC20 tokens for creating a dual-token mechanism that enhances drone service reliability through authentic feedback.

\subsection{Challenges in Traditional Review Systems and Potential Mitigations}
The advent of the IoDT brings to fore the critical challenge of ensuring genuine feedback in a landscape susceptible to fraudulent reviews. Studies have highlighted various biases and potential frauds that plague traditional feedback systems \cite{tadelis2015economics,9490665,cai2016fraud,9678955,9807355}. These encompass a range of deceitful practices from bad-mouthing to ballot-stuffing, necessitating robust mechanisms to mitigate such vulnerabilities.

Mitigation strategies span both incentive-based and non-incentive-based models, with blockchain technology emerging as a key player in non-incentive approaches. These models aim to leverage technological advancements to authenticate reviews without direct rewards \cite{ramachandiran1using,dennis2016rep,azad2018privbox,park2018smart}. However, while promising, these models often exhibit limitations in their current forms, from scalability issues to unaddressed security vulnerabilities.

\subsection{Comparative Analysis}
Our DDRM framework, drawing from the insights gained through the examination of existing systems, introduces an integrated solution that not only addresses the authenticity of drone service reviews but also tackles inherent scalability and security challenges. As depicted in Table \ref{table:comparison}, DDRM stands out for its comprehensive approach to mitigating prevalent attacks in review systems, backed by a practical implementation and a thorough evaluation of operational costs.

In conclusion, the DDRM framework emerges as a cutting-edge solution within the IoDT ecosystem, addressing critical gaps in traditional review systems. By harnessing the power of Ethereum's blockchain and ERC20 tokens, DDRM ensures the reliability and integrity of drone service reviews, paving the way for a more trustworthy IoDT landscape. Future work will focus on enhancing DDRM's scalability and exploring further integration with existing IoDT platforms, reinforcing its utility in diverse operational scenarios.

\begin{figure*}[h]
\centering
\includegraphics[width=\textwidth]{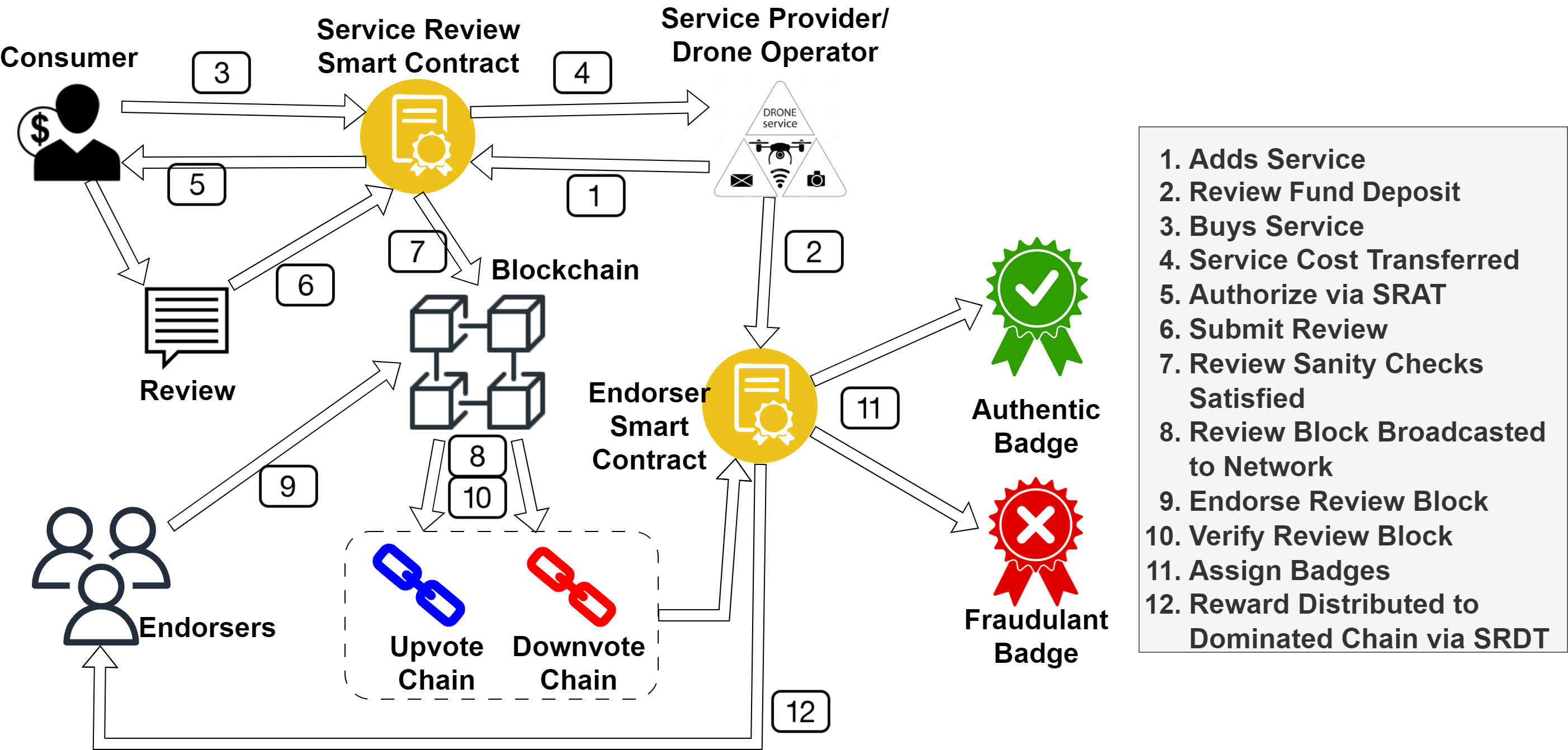}
\caption{Schematic representation of the DDRM framework, illustrating smart contract integration to streamline drone service procurement, review, and endorsement processes within the IoDT.}
\end{figure*}

\section{System Model}
This paper presents the Distributed Drone Reputation Management (DDRM) framework, explicitly designed for enhancing bushfire management through the Internet of Drone Things (IoDT) ecosystem, grounded in the Ethereum blockchain. DDRM innovatively resolves the inherent flaws of traditional drone service review systems with a sophisticated, blockchain-enabled approach. Figure 1 provides a detailed visual representation of DDRM's critical components, emphasizing the role of smart contracts in optimizing the process of drone service procurement, review, and endorsement for bushfire management.

Central to DDRM is its unique capability to integrate drone service procurement with a transparent review mechanism. This integration is crucial for establishing a marketplace where bushfire management authorities, acting as data consumers, can reliably access vital information from drone operators, who serve as data providers. This system ensures the documentation of authentic interactions through the mandatory acquisition of a Service Review Authorization Token (SRAT) by drone service consumers before posting reviews.

To ensure the reliability and authenticity of service reviews in this context, DDRM employs a consortium of peer-to-peer (P2P) endorser nodes. These nodes are instrumental in validating the reviews provided by drone operators, thereby maintaining the integrity and authenticity of feedback crucial for bushfire detection and management.

Further emphasizing our commitment to a trustworthy ecosystem for bushfire management, the introduction of the Drone Reputation Enhancement Token (DRET) acknowledges and rewards drone operators who consistently provide high-quality data, thus enhancing the overall credibility of services within the IoDT for bushfire management.

Subsequent sections explore the DDRM ecosystem in detail, beginning with a comprehensive overview of stakeholders in Section \ref{dsec1}. This includes bushfire management authorities and drone operators, underscoring the marketplace dynamics. Section \ref{dsec2} delves into the operational dynamics and interactions, providing insight into the framework's functionality in bushfire management. Section \ref{dsec3} concentrates on the endorsers, detailing their selection process and pivotal role in ensuring system integrity and the authenticity of data crucial for efficient bushfire management. The procedure for initial endorsement, establishing foundational trust in the data provided, is outlined in Section \ref{dsec4}. Section \ref{dsec5} revisits the endorsers' role in authenticating drone service refunds, ensuring transparency and fairness in the marketplace. Finally, Section \ref{dsec6} discusses DDRM's privacy measures, illustrating the system's commitment to protecting the anonymity and confidentiality of all participants in a secure environment tailored for bushfire management.

\subsection{System Participants in DDRM} \label{dsec1}

Within the DDRM framework, tailored for the Internet of Drone Things (IoDT) ecosystem, we identify four key participant types essential to our unique marketplace for bushfire management: Service Providers (SP), Consumers (C), Reviewers (R), and Endorsers (E).

\begin{enumerate}
    \item \textbf{Service Providers (SP)}: Comprising drones or drone service vendors, these participants offer crucial data collection services for bushfire detection and management, acting as data providers within the IoDT ecosystem.
    
    \item \textbf{Consumers (C)}: This category includes bushfire management authorities that utilize the data and services provided by SPs, sourcing vital information for effective bushfire management.
    
    \item \textbf{Reviewers (R)}: A subgroup of Consumers, these participants provide feedback based on their experiences with drone services, contributing to DDRM's review system. This feedback is pivotal for maintaining service quality and reliability, directly impacting bushfire management outcomes.
    
    \item \textbf{Endorsers (E)}: Emergent from Reviewers, Endorsers are chosen for their consistent honesty and integrity. Automated and randomized selection via smart contracts ensures they play a crucial role in authenticating review veracity, enhancing the marketplace's trustworthiness. Their specific roles are further discussed in Section \ref{dsec3}.
\end{enumerate}

The next section, Section \ref{dsec2}, elaborates on the interactions and activities within DDRM, underpinning our system's operational dynamics.

\subsection{DDRM Operations} \label{dsec2}

DDRM operations, integral to our bushfire management marketplace, incur necessary gas costs in Wei for the execution of smart contracts on the Ethereum blockchain. This underpins the computational efforts within the IoDT ecosystem.

Mandatory registration for Service Providers (SP) and Consumers (C) includes credit/debit card information submission to ensure transactions' accountability. This step prevents multiple registrations per credit/debit card, reinforcing system integrity.

Upon registration, participants receive unique identifiers linked to their Ethereum address(es), facilitating activity tracking within DDRM.

Service Providers (SP) have the capability to introduce new services, modify existing offerings, or withdraw them, pivotal for enriching the bushfire management data repository. Listing a new service \(S\) involves a specific gas cost (\(GC_{\text{AddS}}\)), directly debited from the SP's account. A lack of sufficient funds halts this process. Moreover, to encourage robust review participation, 1 Ether from the SP's account is allocated to the ReviewFund(\(S\)), subsidizing review gas costs for Consumers (C) and ensuring a dynamic feedback loop.

Service registration adjustments are captured in the following equations:

\begin{equation}
\text{Account}(SP) = \text{Account}(SP) - [GC_{\text{AddS}} + 1 \text{Ether}]
\end{equation}

\begin{equation}
\text{ReviewFund}(S) = \text{ReviewFund}(S) + 1 \text{Ether}
\end{equation}

In the DDRM framework, essential for creating a marketplace where bushfire management authorities can obtain crucial information, Consumers (\(C\))—representing these authorities—are enabled to procure drone services and subsequently submit reviews. Upon enlisting a service \(S\), a Consumer (\(C\)) can avail of it through their Account(\(C\)), ensuring they maintain a balance exceeding the service cost (\(sCost\)).

The process debits the gas expenditure \(GC_{\text{BuyS}}\) and \(sCost\) from the Consumer's Account(\(C\)). This transaction awards an ERC20 token, the SRAT, to \(C\), marking their eligibility to review, which is vital for enhancing bushfire management strategies by ensuring the collection of quality data.

\begin{equation}
\text{Account}(C) = \text{Account}(C) - [GC_{\text{BuyS}} + sCost]
\end{equation}

\begin{equation}
\text{SRATBalance}(C) = \text{SRATBalance}(C) + 1
\end{equation}

\begin{equation}
\text{Account}(SP) = \text{Account}(SP) + sCost
\end{equation}

SRAT underwrites the gas costs for review submissions, eliminating financial burdens for Consumers, encouraging them to share their service experiences. This feedback is crucial for bushfire management authorities to assess the effectiveness and reliability of the drone services availed.

The ReviewFund, supported by Service Providers' contributions, ensures a continuous flow of SRAT tokens, facilitating up to a hundred reviews per Ether contributed. This ensures that drone services remain under constant evaluation, promoting transparency and reliability in the data crucial for bushfire detection and management.

Service Providers are incentivized to replenish the ReviewFund to maintain the visibility and operability of their services within this marketplace, directly impacting the efficiency of bushfire management efforts.

Algorithm 1 outlines the review process within DDRM, ensuring that only services availed by legitimate Consumers, such as bushfire management authorities, are reviewed. This procedure ensures that reviews are genuine and reflective of the service quality, contributing significantly to the ongoing improvement of bushfire management strategies.

\begin{algorithm}[t]
\caption{Review Drone Service in DDRM}
\label{alg:review_service}
\begin{algorithmic}[1]
\Require \textit{ServiceID} is the unique ID of the drone service to review.
\Require \textit{userID} is the unique ID of the reviewer.
\Ensure Drone service review.

\Procedure{ReviewService}{ServiceID}
\If{user purchased the drone service with ServiceID \textbf{and} holds an unexpired SRAT token}
    \State Allow user to submit review.
    \State Add user's review to blockchain.
    \State Append userID to the roster of that drone service reviewers.
    \State Utilize SRAT token to underwrite review submission gas cost.
    \State Burn the token.
\Else 
    \State Deny permission to review the drone service.
\EndIf
\EndProcedure
\end{algorithmic}
\end{algorithm}

\subsection{P2P Randomized Consortium Network of Endorsers in DDRM} \label{dsec3}

Within DDRM, crucial for bridging drone operators with bushfire management authorities, each service review undergoes a rigorous endorsement process by endorser nodes, based on their direct experiences with the drone services in question. These endorsements, manifesting as upvotes or downvotes, directly impact the credibility of reviews within our marketplace, distinguishing genuine feedback from potentially misleading information.

Endorsers validate reviews by granting upvotes for feedback that aligns with their experience, signaling authenticity and reliability, or downvotes for reviews perceived as misleading, ensuring only credible data supports bushfire management decisions.

To further validate the integrity of reviews within DDRM, reviews are awarded badges based on endorser consensus:

\begin{itemize}
    \item \textbf{Authentic Badge}: Assigned to reviews that reflect the collective endorsement experience, indicating a review's reliability in accurately representing a drone service's effectiveness in bushfire management scenarios.
    \item \textbf{Fraudulent Badge}: Designated for reviews that diverge from the general endorser experience, suggesting a review may misrepresent the service's utility in managing bushfire detection and response efforts.
\end{itemize}

\begin{algorithm}[t]
\caption{Selection and Reward Mechanism for Endorsers}
\label{alg:endorsers_updated}
\begin{algorithmic}[1]
\Require Collection of reviews: \textit{ReviewsToEndorse}
\Ensure List of potential endorsers: \textit{EndorserCandidates}
\Ensure Chosen endorsers for endorsement: \textit{SelectedEndorsers}
\Ensure Reviewers to penalize: \textit{PenalizeCandidates}
        
\Procedure{EndorserSelection}{addr(R), ServiceID}
\ForAll{reviews in \textit{ReviewsToEndorse}}
\If{review has more upvotes than downvotes}
\State Grant \textbf{Authentic Badge} to review.
\State Add reviewer's address to \textit{EndorserCandidates}.
\Else
\State Assign \textbf{Fraudulent Badge} to review.
\State Accumulate reviewer's address in \textit{PenalizeCandidates}.
\EndIf
\EndFor

\State Clear the current \textit{SelectedEndorsers} list.
\ForAll{reviewers in \textit{EndorserCandidates}}
\State Augment the \textit{SelectedEndorsers} list.
\State Distribute SRDT to eligible reviewers.
\EndFor

\ForAll{reviewers in \textit{PenalizeCandidates}}
\If{total fraudulent badges surpass threshold}
\State Exclude reviewer from DDRM framework.
\EndIf
\EndFor
\EndProcedure
\end{algorithmic}
\end{algorithm}

Algorithm 2 articulates the steps undertaken by DDRM to ensure that endorsements reflect genuine experiences, crucial for maintaining the accuracy and reliability of the data upon which bushfire management decisions are based. The algorithm emphasizes the continuous cycle of endorsement, validation, and reward, reinforcing the DDRM framework's role in fostering a trustworthy environment for the exchange of critical service reviews relevant to bushfire management.

Algorithm 3 outlines the process by which reviewers, elevated to the status of endorsers within the \textit{SelectedEndorsers} group, contribute to the review endorsement mechanism in DDRM, a pivotal component in our marketplace for bushfire management. This mechanism is integral for ensuring the credibility of drone service reviews, which, in turn, influences bushfire management decisions.

To participate in the endorsement process:

\begin{itemize}
    \item Endorsers must be listed in the \textit{SelectedEndorsers}, affirming their role in providing trustworthy feedback.
    \item They need to hold a valid, unexpired Service Review Discount Token (SRDT) for the drone service being reviewed, ensuring a direct connection to the service evaluated.
\end{itemize}

Upon meeting these prerequisites, endorsers can cast their vote on reviews, contributing to the assessment of drone services used in bushfire management. The endorsement, encapsulated as either an upvote or downvote, is recorded alongside the endorser's and service's identifiers, guaranteeing transparency and preventing duplicate endorsements for a single review. Following their endorsement, the SRDT is consumed, emphasizing its one-time use and maintaining the system's integrity.

\begin{algorithm}[t]
\caption{Endorsement Mechanism for Reviews}
\label{alg:review_endorsement_updated}
\begin{algorithmic}[1]
\Require Collected reviews: \textit{ReviewsToConsider}
\Ensure Endorsement annotations: \textit{EndorsementAnnotations}
\Ensure Chosen endorsers: \textit{ChosenEndorsers}
        
\Procedure{EndorsementProcedure}{ReviewsToConsider}
\ForAll{reviews in \textit{ReviewsToConsider}}
\If{Endorser belongs to \textit{ChosenEndorsers} \textbf{and} holds valid, unexpired \textit{SRDT}}
\If{Endorser casts affirmative endorsement}
\State Increase upvotes count of review.
\State \textit{EndorsementAnnotations} appends Endorser's identifier and \textit{ServiceID}.
\State Consume the \textit{SRDT}.
\Else
\State Elevate downvotes count of review.
\State \textit{EndorsementAnnotations} appends Endorser's identifier and \textit{ServiceID}.
\State Consume the \textit{SRDT}.
\EndIf
\Else
\State Endorser denied the endorsement privilege.
\EndIf
\EndFor
\EndProcedure
\end{algorithmic}
\end{algorithm}

\subsection{Bootstrapping Endorser Nodes in DDRM} \label{dsec4}

For DDRM, initially endorsing service reviews—crucial for evaluating drone services in bushfire management—is vital. To bootstrap this process within the IoDT ecosystem, the Endorser Smart Contract autonomously and randomly selects a predefined number of reviewers from the earliest service evaluations to serve as endorsers. This selection not only kickstarts the endorsement mechanism but also rewards these newly appointed endorsers with Service Review Discount Tokens (SRDT), linking them to specific drone services via a unique ServiceID. This enables them to access services at reduced rates and to contribute their endorsements, thereby enhancing the reliability of drone services for bushfire management. Future endorser selections follow the process outlined in Section \ref{dsec3}, ensuring a dynamic and effective endorsement cycle.

\subsection{Endorsers and the Refund Mechanism in DDRM} \label{dsec5}

Refunds become relevant when Consumers, including bushfire management authorities, find a drone service unsatisfactory. DDRM addresses this by involving previously selected endorsers, who have firsthand experience with the services, in the review of refund claims. This ensures that refund requests are scrutinized by those with relevant service experience, maintaining the integrity of the marketplace and preventing unfounded claims from undermining the credibility of drone services essential for bushfire management.

\subsection{Participant Privacy in DDRM} \label{dsec6}

In DDRM, participant privacy is paramount, especially within the IoDT ecosystem where data sensitivity is high, such as in bushfire management scenarios. Employing Ethereum's blockchain technology, DDRM ensures participant anonymity through the use of unique public-private key pairs. This system masks participant identities behind Ethereum addresses, securing their activities within the DDRM framework from external scrutiny. This privacy safeguard is critical, allowing for the secure exchange of information and services crucial for managing and mitigating bushfires effectively.

\section{Analysis and Discussion}

This section embarks on a comprehensive exploration of the DDRM framework, as it pertains to the Internet of Drone Things (IoDT) ecosystem, with a particular emphasis on its application in creating a marketplace for bushfire management. Starting with Section \ref{d51}, we dissect the adversary model by delineating distinct rater categories within DDRM and identifying possible rating frauds that could compromise the integrity of service reviews vital for bushfire management. In Section \ref{d52}, leveraging the adversary model insights, we assess DDRM's robustness in thwarting malicious actions and preserving the credibility of service reviews. Concluding our examination in Section \ref{d53}, we turn our attention to DDRM's performance metrics, including a detailed analysis of the computational overhead, depicted through gas costs, associated with executing transactions and managing smart contracts.

\subsection{Adversary Model in DDRM} \label{d51}

In detailing the adversary model for DDRM, tailored for the IoDT ecosystem and its application in bushfire management, we undertake three crucial steps:

\begin{enumerate}
    \item Identifying rating frauds within DDRM that adversaries might exploit to affect the reliability of drone services for bushfire management.
    \item Categorizing rater profiles participating in DDRM, with a focus on their impact on bushfire management services.
    \item Mapping identified rating frauds to rater categories to pinpoint potential threats and their implications for bushfire management effectiveness.
\end{enumerate}

\subsubsection{Rating Frauds in DDRM} \label{d511}
We delve into rating frauds within DDRM, focusing on how such deceptions could impair drone services' evaluation for bushfire management. Rating frauds, primarily bad-mouthing and ballot-stuffing, could manifest in various forms\cite{dellarocas2000immunizing, irissappane2012towards, cai2016fraud, dellarocas2003digitization, fradkin2015bias, voss2004privacy}:

\begin{itemize}
    \item \textbf{Collusion Attacks:} Stakeholders might conspire to manipulate reviews, directly impacting the selection and reliability of drone services for bushfire management. Examples include service providers incentivizing consumers to modify critical feedback or colluding with external entities to fabricate reviews.
    
    \item \textbf{Constant Attacks:} Targeted manipulation of specific drone services with false reviews to artificially influence their standing within the bushfire management marketplace.
    
    \item \textbf{Whitewashing Attacks:} Perpetrators authoring deceitful reviews and then erasing traces of their activities to avoid detection and accountability within the DDRM system.
    
    \item \textbf{Sybil Attacks:} The creation of multiple fake identities to flood DDRM with bogus reviews, affecting the accuracy of service evaluations crucial for bushfire management.
    
    \item \textbf{51\% Majority Attack:} Gaining control over a majority of the endorser nodes to manipulate the review validation process, potentially skewing the evaluation of drone services for bushfire management.
\end{itemize}

\subsubsection{Association of Rater Categories with Rating Frauds in DDRM} \label{d512}
In our DDRM framework, crucial for ensuring the reliability of drone services for bushfire management, understanding the motives behind reviews is essential. Here, we classify raters based on their intentions and outline the potential rating frauds they may engage in:

\begin{enumerate}
    \item \textbf{Happy Honest Raters:} These raters sincerely appreciate the drone services used in bushfire management and provide positive feedback based on genuine experiences.
    \textbf{Potential Rating Frauds:} Typically immune to engaging in frauds, their authentic reviews are vital for validating the effectiveness of drone services in bushfire scenarios.

    \item \textbf{Unhappy Honest Raters:} These individuals offer genuine negative feedback based on unsatisfactory experiences with drone services.
    \textbf{Potential Rating Frauds:} Their critiques, while honest, could be targeted for suppression or alteration through collusion, risking the overshadowing of legitimate concerns critical for improving bushfire management services.

    \item \textbf{Happy Dishonest Raters:} This group artificially inflates the reputation of certain drone services with false positive reviews, a practice detrimental to the trustworthiness of services pivotal for bushfire detection.
    \textbf{Potential Rating Frauds:} They are prone to engage in a spectrum of deceptive practices, from creating fake accounts (Sybil attacks) to collusion, potentially skewing the perceived reliability of drone services.

    \item \textbf{Unhappy Dishonest Raters:} These raters aim to undermine the reputation of drone services through malicious negative reviews.
    \textbf{Potential Rating Frauds:} Employing tactics similar to their Happy Dishonest counterparts, their actions can significantly distort the landscape of trusted drone services, impacting the selection of effective solutions for bushfire management.
\end{enumerate}

By discerning these rater categories and their potential to commit rating frauds, DDRM aims to safeguard the integrity of reviews critical for bushfire management, ensuring that the most reliable and effective drone services are highlighted and utilized.

\subsection{Security Analysis of DDRM Against Adversary Model} \label{d52}
This subsection evaluates DDRM's defense mechanisms against the adversary model previously outlined, emphasizing its role in securing drone service reviews vital for bushfire management. Table~\ref{table:defense} presents the defense strategies of DDRM for each rater category and their potential attacks.

\begin{table*}[h]
\centering
\caption{Security analysis of DDRM against potential rating attacks by rater categories, emphasizing the context of bushfire management.}
\label{table:defense}
\begin{tabularx}{\textwidth}{p{1in} X X}
\hline
\textbf{Rater Category} & \textbf{Attack Scenarios} & \textbf{DDRM Defense} \\
\hline
Happy Honest & Not applicable, as their feedback stems from genuine, positive experiences with drone services. & — \\

Unhappy Honest & Potential collusion to modify or remove genuine negative reviews crucial for assessing drone service efficacy in bushfire scenarios. & Blockchain's data immutability ensures reviews remain unchanged and permanently recorded, maintaining review integrity. \\

Happy Dishonest & Engaging in fraudulent practices to artificially boost service ratings, misleading bushfire management decisions. & Limiting reviews to one per user and authenticating via tokens reduces fraudulent reviews, while peer-to-peer endorsement verifies review authenticity. \\

Unhappy Dishonest & Attempting to undermine drone services by spreading false negative reviews, potentially affecting bushfire management strategies negatively. & Enforcing a stringent review policy and utilizing peer-to-peer endorsement to scrutinize and validate each review's authenticity. \\
\hline
\end{tabularx}
\end{table*}

DDRM's architecture, integrating service acquisition with review mechanisms, ensures that only genuine users can review, enhancing the feedback's authenticity for bushfire management purposes. The framework's decentralized nature prevents manipulation by central authorities or service providers, with blockchain technology ensuring that once recorded, reviews are immutable and tamper-proof.

Moreover, DDRM values all feedback, enabling a comprehensive evaluation of drone services. Through its innovative dual-token mechanism, reviews are validated by a network of trustworthy endorsers, culminating in a credibility assessment that supports bushfire management authorities in making informed decisions.

Subsequent analyses will further explore DDRM's resilience against various vulnerabilities and attacks, reinforcing its capacity to safeguard the integrity of crucial data for effective bushfire management.

\subsubsection{Collusion Attacks in DDRM}

DDRM mitigates collusion risks—whether between users and drone service providers or between providers and centralized entities—crucial for maintaining trustworthy drone service reviews for bushfire management. The decentralized blockchain foundation of DDRM eliminates central manipulation risks, ensuring that all reviews, once recorded, remain immutable and tamper-proof, reflecting genuine user experiences with drone services.

\subsubsection*{Multi-Identity Collusion}

DDRM's structure effectively deters multi-identity collusion aimed at manipulating reviews within the bushfire management context. The economic impracticality of such attacks, coupled with the P2P endorser model, ensures that fabricated feedback is quickly identified and penalized. This system discourages attempts to undermine competitors through deceitful means, as it would inadvertently contribute to their revenue. Additionally, the unpredictability in endorser node selection and the risk of penalties for fraudulent activities further secure the review process against manipulation.

\subsubsection*{Ballot-stuffing}

Addressing ballot-stuffing, DDRM enforces economic and procedural barriers that deter service providers from engaging in this form of attack. Contributions to the review fund and irrevocable costs tied to SRAT tokens create financial disincentives against generating fake reviews. The endorsement process, requiring validation by impartial endorsers, adds a critical layer of defense, ensuring that only authentic reviews influence the assessment of drone services for bushfire management.

Furthermore, the DRET token mechanism restricts endorsers to review only those services associated with their tokens, preventing any potential bias or manipulation from endorsers who might have been covertly influenced by adversaries. This structured approach enhances DDRM's robustness against fraudulent activities, safeguarding the credibility of drone services essential for effective bushfire detection and management.

\subsubsection{Constant Attacks in DDRM}

DDRM's structure necessitates actual service usage before review, with SRAT tokens ensuring reviewers have genuinely availed the drone services they comment on. This, coupled with the rule that each review can only be endorsed once by an endorser, forms a robust defense against constant attacks, protecting the credibility of drone service reviews essential for bushfire management within the IoDT ecosystem.

\subsubsection{Whitewashing Attacks in DDRM}

The DDRM framework's rigorous registration process, involving credit/debit card details, prevents participants from easily discarding their history to circumvent accountability. This measure ensures the continuity of participant reputations, safeguarding the review ecosystem against whitewashing tactics that could undermine the trust in drone services for bushfire management.

\subsubsection{Sybil Attacks in DDRM}

DDRM mitigates the risk of Sybil attacks through a unique registration procedure, linking participants' identities to their financial credentials and Ethereum addresses. While not entirely eliminating the possibility of Sybil attacks, DDRM's design significantly reduces their likelihood and effectiveness, ensuring the integrity of the drone service review process for bushfire management.

\subsubsection{51\% Majority Attack in DDRM}

Recognizing the potential threat of a 51\% majority attack, DDRM acknowledges this inherent blockchain vulnerability. Future enhancements will focus on integrating advanced security measures to further safeguard the review validation process, crucial for maintaining the reliability of drone services in bushfire management.

\subsubsection{False Refund Claims in DDRM}

Addressing the issue of false refund claims, DDRM employs a smart contract to select endorsers randomly for evaluating refund requests. This mechanism leverages collective experiences to ascertain the authenticity of claims, preventing fraudulent attempts that could adversely affect the assessment and selection of drone services for bushfire management.

\subsection{Implementation and Performance Analysis} \label{d53}
This section assesses the DDRM framework's effectiveness and efficiency, emphasizing its application in bushfire management through smart contract implementations, focusing on gas cost implications and security against potential vulnerabilities.

\subsubsection{Gas Cost Analysis}
Our DDRM prototype was deployed on the Ethereum Ropsten Test Network for empirical analysis, using test ethers to evaluate gas costs associated with DDRM operations, crucial for determining the framework's efficiency in bushfire management scenarios.

In DDRM, operations such as adding services, requesting services, submitting reviews, and endorsing reviews, all necessitate gas, a proxy for the computational resources required. The notable gas consumption for each operation reflects on the DDRM's operational speed and efficiency, vital for timely bushfire detection and management:
\begin{itemize}
    \item \textit{Add Service} by drone service providers consumes gas for updating service data.
    \item \textit{Request Service} by Consumers results in gas deduction for accessing services.
    \item Submitting a review involves gas costs, mitigated by the SRAT provided to the Consumer.
    \item Endorsing reviews also incurs gas costs, deducted from the endorser's account.
\end{itemize}
Smart contracts, especially those modifying on-chain data, are computationally demanding, influencing gas usage.

Table \ref{DDRMGas} presents the gas costs, in both Ether and USD, for DDRM's operations, with \textit{Add Service} exhibiting higher gas usage due to its extensive contract interactions. Despite this, the benefits of enhanced security and trust in the bushfire management context justify the gas costs, underscoring DDRM's cost-effectiveness.

Evaluations were conducted with a 2.9 Gwei gas price, ensuring efficient DDRM contract execution. The absence of similar IoDT systems makes direct comparisons challenging; however, DDRM's operational costs are considered modest, promoting its adoption without significant financial strain on service providers or consumers in bushfire management applications.

\begin{table*}[h]
\centering \caption{Gas cost for the various DDRM modules in the context of bushfire management.} \label{DDRMGas}
\begin{tabular}{|l|l|r|r|r|r|r|}
\hline
\textbf{Function Caller} & \textbf{Function Name} & \textbf{Gas Limit (Units)} & \textbf{Gas Used (Units)} & \textbf{Gas Price (Gwei)} & \textbf{Total} & \textbf{Total (USD)} \\
\hline
Drone Service Provider & Add Service & 272456 & 182304 & 2.9 & 0.000529 & 0.839 \\
\hline
Consumer & Request Service & 99872 & 63789 & 2.9 & 0.000185 & 0.293 \\
\hline
Endorser & Endorse Review & 106754 & 86532 & 2.9 & 0.000251 & 0.398 \\
\hline
\end{tabular}
\end{table*}

\subsubsection{Smart Contract Analysis}

The DDRM framework, proposed for the Internet of Drone Things (IoDT) ecosystem, heavily relies on Ethereum's smart contracts to execute its operations. Ensuring the security and integrity of these contracts is paramount for maintaining trust within the ecosystem. Given the immutability of deployed smart contracts and their transparency to all blockchain users, potential vulnerabilities in the code can become a point of exploitation, reminiscent of incidents like the DAO attack \cite{chen2020survey}. 

In the context of IoDT, where drones operate in dynamic environments and often carry out critical tasks, the security of smart contracts becomes even more pivotal. Any vulnerability can not only disrupt the operations but also compromise the trust and reputation management mechanism of the DDRM framework. Consequently, before deploying any smart contract on the Ethereum mainnet, a comprehensive analysis is essential to identify and rectify potential flaws, ensuring the robustness of our proposed solution.

To ensure the robustness of the smart contracts within our DDRM framework, we utilized SmartCheck \cite{tikhomirov2018smartcheck}, a static analysis tool designed for the Solidity language. This tool is equipped to analyze approximately 75 checks that span across three main vulnerability classes: blockchain, language, and model. Further details about the vulnerability classifications can be found in the official SmartDec Github repository \footnote{https://github.com/smartdec}.

Upon an in-depth analysis of our smart contracts, encompassing approximately 350 lines of code, the tool SmartCheck identified a set of vulnerabilities. One prominent vulnerability pertains to storage access. This issue suggests the potential of an overlap attack due to direct access to storage slots, which could affect certain state variables. Another identified vulnerability is linked to gas limitations. This particular concern arises because of the increased gas cost that can be incurred when evaluating state variables within specific loop conditions. However, it is of paramount importance to underscore that, despite these vulnerabilities, our smart contracts maintain robust security, guarding against a plethora of conceivable exploits and vulnerabilities.

In the DDRM framework, the bedrock of our architecture relies on Ethereum smart contracts. Given the immutable nature of blockchain, and the visibility of smart contracts to all network participants, it becomes pivotal to ensure that these contracts are devoid of vulnerabilities. Any oversight could expose the system to potential attacks, reminiscent of the DAO debacle \cite{chen2020survey}. 

To assuage these concerns, we employed SmartCheck \cite{tikhomirov2018smartcheck} for a thorough static analysis of our smart contracts. This tool, tailored for Solidity, is adept at identifying vulnerabilities spread across three core categories: blockchain, language, and model.

\begin{table*}[h!]
\centering
\caption{SmartCheck's detection of vulnerabilities in DDRM Smart Contracts
}
\begin{tabular}{|c|c|c|}
\hline
\textbf{Vulnerability Class} & \textbf{Vulnerability Group} & \textbf{Detection in DDRM Smart Contracts} \\
\hline
Blockchain & Ether transfer & No \\
\cline{2-3}
& Message Structure & No \\
\cline{2-3}
& Gas Limitations & Yes \\
\cline{2-3}
& Contract Interaction & No \\
\cline{2-3}
& Block content manipulation & No \\
\hline
Language & Internal control flow & No \\
\cline{2-3}
& Storage access & Yes \\
\cline{2-3}
& Arithmetic & No \\
\hline
Model & Economy & No \\
\cline{2-3}
& Privacy & No \\
\cline{2-3}
& Trust & No \\
\cline{2-3}
& Authorization & No \\
\hline
\end{tabular}
\end{table*}

From the analysis, vulnerabilities linked with gas limitations and storage access were identified. It's imperative to note that the DDRM smart contracts are predominantly secure, minimizing the risk of potential exploits.

\subsection{Discussion} \label{d54}
The DDRM framework marks a significant advancement in establishing trust within the IoDT ecosystem, crucial for applications like bushfire management. While DDRM provides a robust solution for validating drone service reliability and behavior, there are areas for refinement and potential challenges to address.

\subsubsection{Updating Drone Service Reviews}
Our initial DDRM design permits only one review per drone service to simplify operations. However, to accommodate evolving interactions with drone services, especially in dynamic scenarios like bushfire detection, DDRM could enable multiple reviews per entity. This extension would allow updates based on new experiences, incorporating a review chain concept to ensure flexibility while deterring spam through review limitations.

\subsubsection{Scalability Concerns}
As drones become increasingly prevalent in critical sectors like bushfire management, addressing DDRM's scalability is paramount. The Ethereum blockchain's transactional capacity may be challenged by the surge in IoDT activities. Exploring advanced blockchain scalability solutions could ensure DDRM's performance meets the demands of extensive drone operations without compromising its core functionalities.

\subsubsection{Integration with IoDT Platforms}
For DDRM to effectively contribute to bushfire management, its integration into broader IoDT platforms is essential. DDRM can function as both an independent module for managing drone reputations and a complementary component within existing IoDT infrastructures, enhancing decentralized reputation management capabilities.

\subsubsection{Reputation Enhancement Fund}
The current static allocation for DDRM's reputation enhancement fund may not align with the dynamic economic realities of IoDT operations. Adopting an adaptive model, where a percentage of the drone service's cost contributes to the fund, could offer a more economically viable approach. This method would dynamically adjust DRET token issuance based on service prices and computational costs, aligning more closely with the operational dynamics of bushfire management.

In conclusion, while DDRM presents a promising solution for enhancing trust and reliability in drone services crucial for bushfire management, future work will focus on addressing scalability, review update mechanisms, economic viability, and seamless integration with existing IoDT platforms. These enhancements aim to solidify DDRM's foundation as a pivotal component in the trustworthy and efficient utilization of drone technologies for bushfire detection and response.

\section{Conclusion and Future Work}
This paper has presented the DDRM framework, tailored for enhancing trust and reliability in the Internet of Drone Things (IoDT), with a special focus on bushfire management applications. The DDRM framework, built upon the Ethereum blockchain, introduces a novel approach to ensuring service reliability and validating drone behavior, crucial aspects in the dynamic IoDT landscape, especially in bushfire detection and management scenarios. The implementation of a dual-token mechanism, including the SRAT and the DRET, alongside a peer-to-peer network of endorser nodes, facilitates a robust environment for verifying drone service reviews.
A thorough security analysis underscored DDRM's resilience against potential threats, validating its effectiveness through deployment on test networks. The operational cost analysis demonstrated DDRM's economic feasibility, marking it as a practical solution for IoDT stakeholders, particularly those involved in bushfire management.

Looking ahead, the exploration of scalability options through private blockchain networks and the optimization of token allocation strategies stand out as pivotal areas for development. These enhancements aim to ensure DDRM's sustainability and to refine incentives for promoting honest drone behavior, thereby improving the framework's utility in bushfire management. Future research will also delve into integrating advanced security measures to fortify DDRM against evolving threats and to expand its application scope within the IoDT ecosystem, ultimately contributing to more effective and reliable bushfire detection and response efforts.

\bibliographystyle{ieeetr}
\bibliography{refs}

\begin{thebibliography}{10}

\bibitem{akram2023chained}
J.~Akram, M.~Umair, R.~H. Jhaveri, M.~N. Riaz, H.~Chi, and S.~Malebary, ``Chained-drones: Blockchain-based privacy-preserving framework for secure and intelligent service provisioning in internet of drone things,'' {\em Computers and Electrical Engineering}, vol.~110, p.~108772, 2023.

\bibitem{akram2022bc}
J.~Akram, A.~Akram, R.~H. Jhaveri, M.~Alazab, and H.~Chi, ``Bc-iodt: blockchain-based framework for authentication in internet of drone things,'' in {\em Proceedings of the 5th international ACM mobicom workshop on drone assisted wireless communications for 5G and beyond}, pp.~115--120, 2022.

\bibitem{munawar2022framework}
H.~S. Munawar, Z.~Gharineiat, and S.~Imran~Khan, ``A framework for burnt area mapping and evacuation problem using aerial imagery analysis,'' {\em Fire}, vol.~5, no.~4, p.~122, 2022.

\bibitem{munawar2022civil}
H.~S. Munawar, F.~Ullah, D.~Shahzad, A.~Heravi, and S.~Qayyum, ``Civil infrastructure damage and corrosion detection: An application of machine learning,'' {\em Buildings}, vol.~12, no.~2, p.~156, 2022.

\bibitem{munawar2023drone}
H.~S. Munawar, S.~I. Khan, F.~Ullah, and B.~J. Choi, ``Drone-as-a-service (daas) for covid-19 self-testing kits delivery in smart healthcare setups: a technological perspective,'' {\em ICT Express}, vol.~9, no.~4, pp.~748--753, 2023.

\bibitem{tahir2022automatic}
A.~Tahir, H.~S. Munawar, M.~Adil, S.~Ali, A.~Z. Kouzani, and M.~P. Mahmud, ``Automatic target detection from satellite imagery using machine learning,'' {\em Sensors}, vol.~22, no.~3, p.~1147, 2022.

\bibitem{akram2024DroneSSL}
J.~Akram, A.~Anaissi, W.~Othman, A.~Alabdulatif, and A.~Akram, ``Dronessl: Self-supervised multimodal anomaly detection in internet of drone things,'' {\em IEEE Transactions on Consumer Electronics}, vol.~70, no.~1, pp.~4287--4298, 2024.

\bibitem{lakhdari2020composing}
A.~Lakhdari, A.~Bouguettaya, S.~Mistry, and A.~G. Neiat, ``Composing energy services in a crowdsourced iot environment,'' {\em IEEE Transactions on Services Computing}, vol.~15, no.~3, pp.~1280--1294, 2020.

\bibitem{mukherjee2023isocialdrone}
A.~Mukherjee, N.~Dey, A.~Mondal, D.~De, and R.~G. Crespo, ``isocialdrone: Qos aware mqtt middleware for social internet of drone things in 6g-sdn slice,'' {\em Soft Computing}, vol.~27, no.~8, pp.~5119--5135, 2023.

\bibitem{bahutair2021multi}
M.~Bahutair, A.~Bouguettaya, and A.~G. Neiat, ``Multi-perspective trust management framework for crowdsourced iot services,'' {\em IEEE Transactions on Services Computing}, vol.~15, no.~4, pp.~2396--2409, 2021.

\bibitem{10535995}
J.~Akram, A.~Anaissi, R.~S. Rathore, R.~H. Jhaveri, and A.~Akram, ``Digital twin-driven trust management in open ran-based spatial crowdsourcing drone services,'' {\em IEEE Transactions on Green Communications and Networking}, vol.~8, no.~2, pp.~841--855, 2024.

\bibitem{mukherjee2022disastdrone}
A.~Mukherjee, D.~De, N.~Dey, R.~G. Crespo, and E.~Herrera-Viedma, ``Disastdrone: A disaster aware consumer internet of drone things system in ultra-low latent 6g network,'' {\em IEEE Transactions on Consumer Electronics}, vol.~69, no.~1, pp.~38--48, 2022.

\bibitem{chatterjee2001online}
P.~Chatterjee, ``Online reviews: do consumers use them?,'' 2001.

\bibitem{islam2021blockchain}
A.~Islam, T.~Rahim, M.~Masuduzzaman, and S.~Y. Shin, ``A blockchain-based artificial intelligence-empowered contagious pandemic situation supervision scheme using internet of drone things,'' {\em IEEE Wireless Communications}, vol.~28, no.~4, pp.~166--173, 2021.

\bibitem{10492460}
J.~Akram, A.~Anaissi, R.~S. Rathore, R.~H. Jhaveri, and A.~Akram, ``Galtrust: Generative adverserial learning-based framework for trust management in spatial crowdsourcing drone services,'' {\em IEEE Transactions on Consumer Electronics}, vol.~70, no.~1, pp.~2285--2296, 2024.

\bibitem{mukherjee2020edgedrone}
A.~Mukherjee, N.~Dey, and D.~De, ``Edgedrone: Qos aware mqtt middleware for mobile edge computing in opportunistic internet of drone things,'' {\em Computer Communications}, vol.~152, pp.~93--108, 2020.

\bibitem{9422738}
G.~D. Putra, V.~Dedeoglu, S.~S. Kanhere, R.~Jurdak, and A.~Ignjatovic, ``Trust-based blockchain authorization for iot,'' {\em IEEE Transactions on Network and Service Management}, vol.~18, no.~2, pp.~1646--1658, 2021.

\bibitem{10130620}
S.~Ben~Saad, B.~Brik, and A.~Ksentini, ``Toward securing federated learning against poisoning attacks in zero touch b5g networks,'' {\em IEEE Transactions on Network and Service Management}, vol.~20, no.~2, pp.~1612--1624, 2023.

\bibitem{10547221}
J.~Akram, M.~Aamir, R.~Raut, A.~Anaissi, R.~H. Jhaveri, and A.~Akram, ``Ai-generated content-as-a-service in iomt-based smart homes: Personalizing patient care with human digital twins,'' {\em IEEE Transactions on Consumer Electronics}, pp.~1--1, 2024.

\bibitem{ramachandiran1using}
R.~Ramachandiran, ``Using blockchain technology to improve trust in ecommerce,'' {\em Retrieval Number: 100.1/ijitee. J944108101021 DOI: 10.35940/ijitee. J9441. 08101021 Journal Website: www. ijitee. org Published By: Blue Eyes Intelligence Engineering and Sciences Publication}.

\bibitem{dennis2016rep}
R.~Dennis and G.~Owenson, ``Rep on the roll: a peer to peer reputation system based on a rolling blockchain,'' {\em International Journal for Digital Society}, vol.~7, no.~1, pp.~1123--1134, 2016.

\bibitem{azad2018privbox}
M.~A. Azad, S.~Bag, and F.~Hao, ``Privbox: Verifiable decentralized reputation system for online marketplaces,'' {\em Future Generation Computer Systems}, vol.~89, pp.~44--57, 2018.

\bibitem{park2018smart}
J.-S. Park, T.-Y. Youn, H.-B. Kim, K.-H. Rhee, and S.-U. Shin, ``Smart contract-based review system for an iot data marketplace,'' {\em Sensors}, vol.~18, no.~10, p.~3577, 2018.

\bibitem{carboni2015feedback}
D.~Carboni, ``Feedback based reputation on top of the bitcoin blockchain,'' {\em arXiv preprint arXiv:1502.01504}, 2015.

\bibitem{schaub2016trustless}
A.~Schaub, R.~Bazin, O.~Hasan, and L.~Brunie, ``A trustless privacy-preserving reputation system,'' in {\em ICT Systems Security and Privacy Protection: 31st IFIP TC 11 International Conference, SEC 2016, Ghent, Belgium, May 30-June 1, 2016, Proceedings 31}, pp.~398--411, Springer, 2016.

\bibitem{salah2019blockchain}
K.~Salah, A.~Alfalasi, and M.~Alfalasi, ``A blockchain-based system for online consumer reviews,'' in {\em IEEE INFOCOM 2019-IEEE Conference on Computer Communications Workshops (INFOCOM WKSHPS)}, pp.~853--858, IEEE, 2019.

\bibitem{wood2014ethereum}
G.~Wood {\em et~al.}, ``Ethereum: A secure decentralised generalised transaction ledger,'' {\em Ethereum project yellow paper}, vol.~151, no.~2014, pp.~1--32, 2014.

\bibitem{dannen2017introducing}
C.~Dannen, {\em Introducing Ethereum and solidity}, vol.~1.
\newblock Springer, 2017.

\bibitem{tadelis2015economics}
S.~Tadelis, ``The economics of reputation and feedback systems in e-commerce marketplaces,'' {\em IEEE Internet Computing}, vol.~20, no.~1, pp.~12--19, 2015.

\bibitem{9490665}
M.~Li, L.~Zhu, Z.~Zhang, C.~Lal, M.~Conti, and M.~Alazab, ``Anonymous and verifiable reputation system for e-commerce platforms based on blockchain,'' {\em IEEE Transactions on Network and Service Management}, vol.~18, no.~4, pp.~4434--4449, 2021.

\bibitem{cai2016fraud}
Y.~Cai and D.~Zhu, ``Fraud detections for online businesses: a perspective from blockchain technology,'' {\em Financial Innovation}, vol.~2, pp.~1--10, 2016.

\bibitem{9678955}
L.~D. Tsobdjou, S.~Pierre, and A.~Quintero, ``An online entropy-based ddos flooding attack detection system with dynamic threshold,'' {\em IEEE Transactions on Network and Service Management}, vol.~19, no.~2, pp.~1679--1689, 2022.

\bibitem{9807355}
Z.~Qu, C.~Lyu, and C.-H. Chi, ``Mush: Multi-stimuli hawkes process based sybil attacker detector for user-review social networks,'' {\em IEEE Transactions on Network and Service Management}, vol.~19, no.~4, pp.~4600--4614, 2022.

\bibitem{dellarocas2000immunizing}
C.~Dellarocas, ``Immunizing online reputation reporting systems against unfair ratings and discriminatory behavior,'' in {\em Proceedings of the 2nd ACM Conference on Electronic Commerce}, pp.~150--157, 2000.

\bibitem{irissappane2012towards}
A.~A. Irissappane, S.~Jiang, and J.~Zhang, ``Towards a comprehensive testbed to evaluate the robustness of reputation systems against unfair rating attack.,'' in {\em UMAP Workshops}, vol.~12, 2012.

\bibitem{dellarocas2003digitization}
C.~Dellarocas, ``The digitization of word of mouth: Promise and challenges of online feedback mechanisms,'' {\em Management science}, vol.~49, no.~10, pp.~1407--1424, 2003.

\bibitem{fradkin2015bias}
A.~Fradkin, E.~Grewal, D.~Holtz, and M.~Pearson, ``Bias and reciprocity in online reviews: Evidence from field experiments on airbnb.,'' {\em EC}, vol.~15, pp.~15--19, 2015.

\bibitem{voss2004privacy}
M.~Voss, ``Privacy preserving online reputation systems,'' in {\em Information Security Management, Education and Privacy: IFIP 18th World Computer Congress TC11 19th International Information Security Workshops 22--27 August 2004 Toulouse, France}, pp.~249--264, Springer, 2004.

\bibitem{chen2020survey}
H.~Chen, M.~Pendleton, L.~Njilla, and S.~Xu, ``A survey on ethereum systems security: Vulnerabilities, attacks, and defenses,'' {\em ACM Computing Surveys (CSUR)}, vol.~53, no.~3, pp.~1--43, 2020.

\bibitem{tikhomirov2018smartcheck}
S.~Tikhomirov, E.~Voskresenskaya, I.~Ivanitskiy, R.~Takhaviev, E.~Marchenko, and Y.~Alexandrov, ``Smartcheck: Static analysis of ethereum smart contracts,'' in {\em Proceedings of the 1st international workshop on emerging trends in software engineering for blockchain}, pp.~9--16, 2018.

\end{thebibliography}

\end{document}